# Studying the Magnetic Properties of CoSi Single Crystals[1]

V. N. Narozhnyi* and V. N. Krasnorussky

*Vereshchagin Institute for High Pressure Physics, Russian Academy of Sciences, Troitsk, Moscow oblast, 142190 Russia*
*e-mail: narozhnyivn@gmail.com*

**Abstract**—The magnetic properties of CoSi single crystals have been measured in a range of temperatures $T = 5.5$–$450$ K and magnetic field strengths $H \leq 11$ kOe. A comparison of the results for crystals grown in various laboratories allowed the temperature dependence of magnetic susceptibility $\chi(T) = M(T)/H$ to be determined for a hypothetical "ideal" (free of magnetic impurities and defects) CoSi crystal. The susceptibility of this ideal crystal in the entire temperature range exhibits a diamagnetic character. The $\chi(T)$ value significantly increases in absolute value with decreasing temperature and exhibits saturation at the lowest temperatures studied. For real CoSi crystals of four types, paramagnetic contributions to the susceptibility have been evaluated and nonlinear (with respect to the field) contributions to the magnetization have been separated and taken into account in the calculations of $\chi(T)$.

**DOI:** 10.1134/S106377611305021X

## 1. INTRODUCTION

The intermetallic compound CoSi belongs to a series of transition metal silicides that crystallize in a cubic (B20 type) lattice without a center of inversion. In this series, most detailed investigations have been devoted to manganese silicide MnSi, which is commonly accepted to possess an itinerant (band) character and exhibits a transition to the long-period helimagnetic state at $T < T_N = 28.8$ K.

Cobalt silicide CoSi, in which cobalt ions (like iron ions in FeSi) do not possess magnetic moments, has been studied to a lesser extent. Available published data [1–4] on the magnetic properties of CoSi are rather contradictory.

## 2. EXPERIMENTAL

In this work, we have studied the static magnetic properties of four CoSi single crystals grown in various laboratories, including Ames Lab (United States), Institute of Metal Physics (Russian Academy of Sciences, Yekaterinburg, Russia), and TU Braunschweig (Germany) (two crystals). These samples will be referred to as Ames, Ural, and No. 17 and No. 144, respectively. The crystals were grown by the Bridgman (Ames) and Czochralski method (Ural, No. 17, and No. 144) as described in more detail elsewhere (see, e.g., [5]); they had weights within 130–280 mg.

The magnetization $M$ of CoSi crystals was measured using a vibrating-sample magnetometer (Lake Shore Cryotronix Inc.) equipped with a helium flow cryostat. The measurements were performed in a range of magnetic field strengths ($H \leq 11$ kOe) and temperatures ($T = 5.5$–$450$ K). The $M(H)$ curves were measured at a fixed temperature, and the $M(T)$ curves were measured in various constant fields. Immediately prior to magnetization measurements, the samples were etched in an HCl–HF–HNO$_3$–glycerol (20 : 20 : 20 : 40 v/v) mixture. This pretreatment was very important, since exposure of samples for 24 h in air led to significant changes in their $M(H)$ curves, which exhibited hysteresis even at room temperature.

The sample for magnetic measurements was glued to a silica holder. Since temperature variation at $H = 10$ kOe was accompanied by very small changes in magnetization (from $-650$ to $+300$ μG cm$^3$), it was also necessary to measure and take into account the background signal from the empty holder for each sample. This contribution typically amounted to 260–420 μG cm$^3$ (depending on the sample position determined by its length) in a field of $H = 10$ kOe (for which $M(T)$ measurements are typically performed). In these measurements, the sample center must coincide with that of the field of the magnetometer's electromagnet. For this reason, the holder positions for samples of various lengths were different. To determine the magnetization of a sample, the magnetization of the empty holder was measured in a separate experiment and then subtracted from the magnetization of the sample–holder system. In a standard $M(H)$ measurement cycle, the magnetic field was first varied from zero to $+11$ kOe, then reduced to $-11$ kOe, and eventually increased again to $+11$ kOe. This procedure allowed us to reveal the appearance of irreversibility in $M(H)$, which was related to the ferromagnetic ordering of a certain fraction of magnetic centers at low temperatures.

---

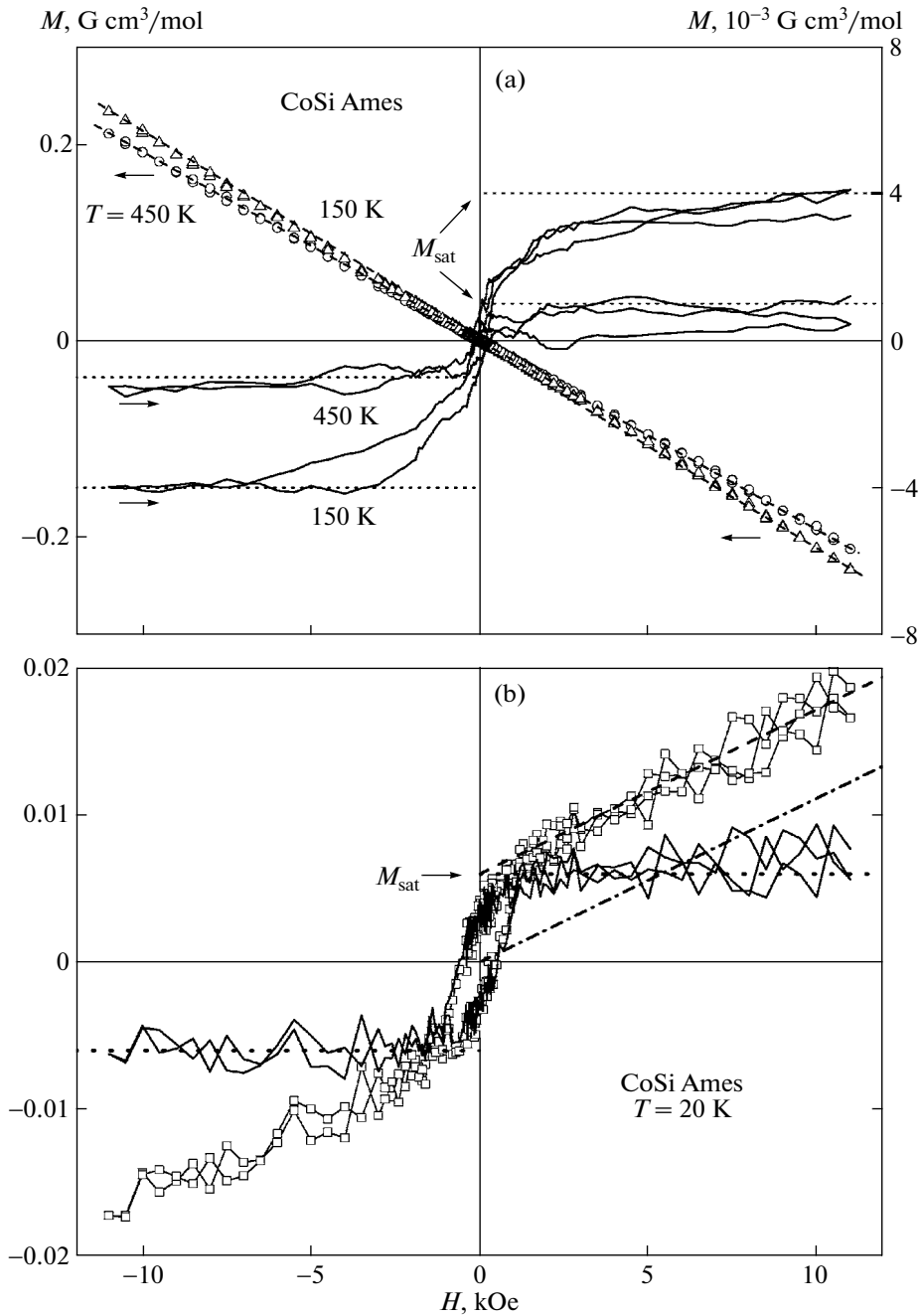

**Fig. 1.** Typical $M(H)$ curves of CoSi single crystal (Ames sample) measured in the (a) diamagnetic [$T = 450$ K (○) and 150 K (△)] and (b) paramagnetic [$T = 20$ K (□)] regions: (a) dashed lines drawn through experimental points illustrate the method of determination of the linear contribution; solid curves show nonlinear contributions at $T = 150$ and 450 K on a greater (right) scale (linear contributions are omitted); horizontal dotted lines show saturation magnetizations $M_{sat}(H)$; (b) separation of $M(H)$ data at $T = 20$ K into linear (dash–dot line) and nonlinear (solid curve) contributions; dashed line illustrates the method of determining the linear $M(H)$ contribution in strong ($H \geq 3$ kOe) fields; the horizontal dotted line shows saturation magnetization $M_{sat}(H)$.

## 3. RESULTS AND DISCUSSION

The results of $M(H)$ measurements showed that these curves were close to linear in the entire temperature range studied for all samples in magnetic fields above approximately 3 kOe. In lower fields, the $M(H)$ curves exhibited some nonlinearities, which were more clearly manifested at low temperatures (Fig. 1). The character of these nonlinearities significantly depended on the measurement temperatures.

At $T \geq 150$ K, the $M(H)$ curves measured for all samples were reversible. A relatively small nonlinear (with respect to the field) contribution to the magneti-zation could be readily determined by subtracting the prevailing linear contribution, which was determined by a conventional method using the results of measurements in strong fields, where the nonlinear contribution exhibited saturation (see Fig. 1). The nonlinear contribution determined using this procedure can be related to the superparamagnetic behavior of a small fraction of magnetic centers present in the studied samples.

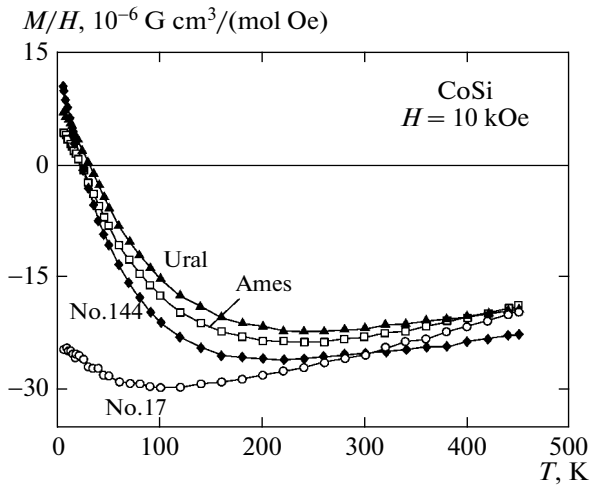

**Fig. 2.** Temperature dependences of magnetic susceptibility $\chi(T) = M(T)/H$ for four CoSi single crystals measured in a field of $H = 10$ kOe. Curves connecting symbols are guides for eye.

When the temperature was reduced to about 100 K, the $M(H)$ curves of all samples except No. 144 showed a hysteresis characteristic of ferromagnetic order, which implied that a small fraction of magnetic centers exhibited the transition to an ordered ferromagnetic state. As the temperature was further reduced, the width of the hysteresis loop significantly increased. Estimations of the concentration of magnetic centers involved in the ferromagnetic ordering gave values on the order of 0.1–1 ppm. For sample No. 144, a hysteresis in the $M(H)$ curves was not observed when the temperature was reduced even to $T = 5.5$ K.

By extrapolating the results of measurements in strong fields to $H = 0$, it is possible to determine the values of saturation magnetization $M_{sat}(T)$ that characterize a nonlinear (with respect to the magnetic field) contribution to the magnetization of samples. These nonlinear $M_{sat}(T)$ corrections were subtracted from the $M(T)$ values measured at $H = 10$ kOe. In this way, we determined the temperature dependences of the magnetization contribution linearly dependent on the field at all temperatures $T$ in the range studied. The $M_{sat}(T)$ corrections for all samples did not exceed 6% of the magnetization measured in a field of 10 kOe at $T = 300$ K.

Figure 2 presents the temperature dependences of magnetic susceptibility $\chi(T) = M(T)/H$ for four CoSi crystals measured in a field of $H = 10$ kOe using the above-described procedure. As can be seen, all samples exhibit a diamagnetic behavior at $T > 40$ K and are characterized by increased magnetization with temperature for $T > 220$ K. To a first approximation, the $M(T)$ curves at $T > 220$ K are close to linear. At low temperatures, the curves of all samples exhibit clearly pronounced paramagnetic "tails." For sample No. 17, the paramagnetic contribution is about ten times as small as those for the other three samples. Indeed, these three samples become paramagnetic at $T < 40$ K, while sample No. 17 remains diamagnetic in the entire temperature range studied. On the other hand, samples No. 17, Ames, and Ural possess close magnetizations at high temperatures (near $T = 450$ K), whereas the magnetization of sample No. 144 is significantly lower. This difference correlates somewhat with the behavior of the nonlinear contribution to the magnetization of sample No 144 in comparison to the other three CoSi crystals (see below).

By comparing the results for various crystals, we determined the temperature dependence of magnetic susceptibility $\chi(T)$ for a hypothetical ideal (free of magnetic impurities and defects) CoSi crystal. This comparison allowed us to eliminate difficulties in separating the paramagnetic contribution to the susceptibility. These difficulties are related to the strong temperature dependence of the diamagnetic contribution at relatively high temperatures.

The procedure used to separate the diamagnetic contribution to $\chi(T)$ was as follows. By subtracting the data for one sample (e.g., No. 17) from those for another sample (e.g., Ames), it is possible in a first approximation to eliminate the temperature-dependent diamagnetic component, reliably determine the paramagnetic contribution, and hence, correctly restore the magnetic susceptibility of the ideal CoSi crystal (see Figs. 3 and 4). In the ideal case, assuming the same properties of the "matrices" in various samples, it is possible to believe that the difference $\chi(T)$ must be well approximated by the Curie–Weiss law as $\chi(T) = C/(T - \theta)$ (without $\chi_0$ term). For real samples prepared in various laboratories, in which the properties of matrices are also slightly different, the experimental data on $\chi(T)$ differences should be approximated using the modified Curie–Weiss relation $\chi(T) = C/(T - \theta) + \chi_0$, where $\chi_0$ is a constant contribution. Note that the $\chi_0$ values for all pairs of crystals were very small, which characterized the closeness of the matrix properties in various samples.

The $\chi(T)$ curves obtained in this way for the ideal CoSi crystal were very close for all pairs of crystals. In the entire working temperature range, the magnetic susceptibility of the hypothetical ideal CoSi crystal possessed a diamagnetic character and its absolute value exhibited significant growth with decreasing temperature. In the region of lowest temperatures, the $\chi(T)$ curve of the ideal CoSi crystal exhibits saturation. This behavior is apparently related to the influence of band effects analogous to those established for FeSi [6].

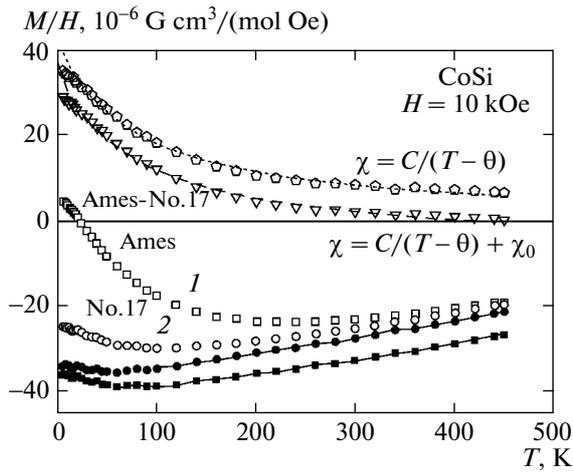

**Fig. 3.** Temperature dependences of magnetic susceptibility $\chi(T) = M(T)/H$ of (□) Ames and (○) No. 17 CoSi single crystals measured in a field of $H = 10$ kOe; (▽) difference Ames–No. 17; (dashed curve) approximation of this difference by the modified Curie–Weiss relation $\chi(T) = C/(T - \Theta) + \chi_0$; (dotted curve) Curie–Weiss relation without $\chi_0$ constant; (◇) separated paramagnetic contribution; (■) $\chi(T)$ of the hypothetical ideal CoSi obtained using experimental data for Ames sample; (●) same for sample No. 17 (solid curves are drawn for illustration).

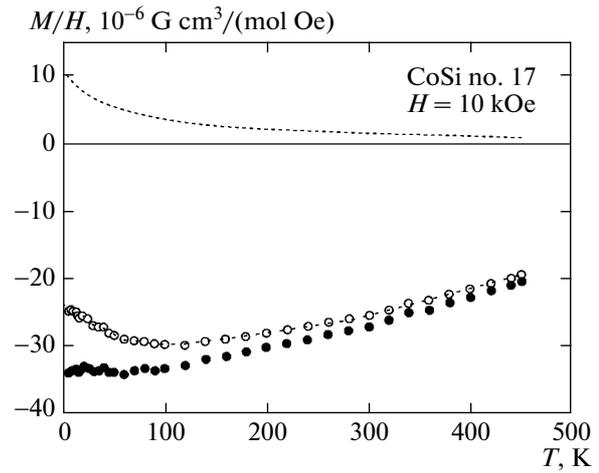

**Fig. 4.** Temperature dependences of magnetic susceptibility $\chi(T) = M(T)/H$ of (○) No. 17 CoSi single crystal measured in a field of $H = 10$ kOe; (dashed curve) separated paramagnetic contribution; (black circles) susceptibility of the hypothetical ideal CoSi (same as depicted by black circles in Fig. 3); (doted curve) sum of contributions plotted by the dashed curve and black circles.

It should be noted that, at $T < T_M \approx 20$ K, all real CoSi samples studied exhibited deviations toward lower magnetization values with respect to the Curie–Weiss type of dependences characteristic of higher temperatures. These deviations were most clearly pronounced for the Ames and Ural samples (see Figs. 2 and 3). It can be assumed that these deviations are related to magnetic ordering of some type (probably with the formation of short-range magnetic order) in some system of magnetic centers, which lead at higher temperatures ($T > 20$ K) to a paramagnetic contribution to $\chi(T)$ that is well described by the modified Curie–Weiss relation $\chi(T) = C/(T - \theta) + \chi_0$.

The nature of the paramagnetic centers responsible for the $\chi(T)$ contribution close to the Curie–Weiss law is unclear. On the one hand, the results of atomic-emission spectroscopy measurements show that the mass content of iron impurity in all CoSi crystals except No. 17 does not exceed 50 ppm, which is insufficient to explain the appearance of paramagnetic tails at low temperatures (see table). The Fe content in Sample No. 17 has been estimated at about 200 ppm, which in principle does not exclude that the low-temperature paramagnetic tails in this case are related to the presence of iron impurity. However, there is no evi-

Magnetic characteristics of four CoSi single crystals

| Sample | Ames | Ural | No. 17 | No. 144 |
|---|---|---|---|---|
| $C$, $10^{-3}$ cm$^3$ K/mol | 3.7 | 4.9 | 0.84 | 2.8 |
| $\theta$, K | −76 | −96 | −76 | −51 |
| $n_{PM}$ (Fe$^{3+}$, $\mu_{eff} = 5.9\mu_B$), ppm | 850 | 1100 | 190 | 640 |
| $n_{PM}$ (Co$^{3+}$, $\mu_{eff} = 4.9\mu_B$), ppm | 1200 | 1600 | 280 | 930 |
| $n_{PM + SPM}$ (Fe$^{3+}$, $\mu = 5\mu_B$), ppm | 0.2 | 0.2 | 0.4 | 0.2 |
| $T_C$, K | 125 | 75 | 75 | <5.5 |
| $T_M$, K | ~20 | ~20 | ~20 | ~20 |

Note: $n_{PM}$ is the concentration of paramagnetic centers in a sample under the assumption that each center possesses an effective magnetic moment $\mu_{eff} = 5.9\mu_B$ (which corresponds to the moment of a free Fe$^{3+}$ ion), $\mu_{eff} = 4.9\mu_B$ (magnetic moment of free Co$^{3+}$ ion); $n_{FM + SPM}$ is the concentration of magnetic centers involved in the ferromagnetic and superparamagnetic behavior under the assumption that each center possesses a magnetic moment projection $\mu = 5\mu_B$ (spin moment of free Fe$^{3+}$ ion, $S = 5/2$); $T_C$ is the Curie temperature; and $T_M$ is the temperature of the onset of deviation from the Curie–Weiss law.

dence that Fe impurity ions in CoSi possess magnetic moments (indeed, iron atoms in the isostructural compound FeSi, as well as cobalt atoms in CoSi, are nonmagnetic, while $Co_{1-x}Fe_xSi$ solid solutions with almost all intermediate Fe concentrations are magnetic; see, e.g., [7]). On the other hand, it was pointed out [2, 8, 9] that, when an iron impurity is introduced into CoSi, magnetic moments can appear on nearest-neighbor Co atoms rather than on impurity atoms. In addition, it should be noted that magnetic moments (per impurity atom) close to those observed in Fe-substituted Co have also been obtained when Co is replaced by Ni, Ru, and Rh. This probably indicates that the magnetic moments that form upon substitution are related to Co rather than to impurity atoms [2]. Alternative possibilities include localization of charge carriers on defects [4] (recently, the localization of carriers on defects was attributed to self-doping due to the polyvalent character of Co and the nonstoichiometry inherent in CoSi [5]). It is also not excluded that magnetic moments form on Co atoms adjacent to some structural defects (e.g., Co or Si vacancies). It was pointed out (although for the Co impurity in FeSi) [10] that an important circumstance in the formation of magnetic centers is the simultaneous presence of Co and Fe atoms at adjacent lattice sites or, in other words, Fe–Co hybridization.

## 4. CONCLUSIONS

Comparison of the results in our experiments with CoSi single crystals grown in various laboratories allowed the temperature dependence of magnetic susceptibility $\chi(T) = M(T)/H$ to be determined for a hypothetical ideal (free of magnetic impurities and defects) CoSi crystal. The susceptibility of this ideal crystal in the entire temperature range possesses a diamagnetic character, shows a significant increase in absolute value with decreasing temperature, and exhibits saturation at the lowest temperatures studied. For real CoSi crystals of four types, the paramagnetic contributions to the susceptibility have been estimated and the nonlinear (with respect to the field) contributions to magnetization have been determined and taken into account in calculating $\chi(T)$. Elucidation of the nature of paramagnetic centers in CoSi crystals requires additional investigations.

## ACKNOWLEDGMENTS

The authors are grateful to S.M. Stishov for kindly providing CoSi samples and helpful discussion of results, and to A.V. Gulyutin for spectrometric analysis of the sample compositions.

*Translated by P. Pozdeev*